\documentclass{emulateapj}
\usepackage{graphicx} 
\usepackage{amsmath}
\usepackage{subfigure}

\shorttitle{Detecting Close-Separation Binary Systems in Synoptic Sky Surveys}
\shortauthors{E. Terziev, N. Law, et al.}

\begin{document}

\title{Millions of Multiples: Detecting and Characterizing Close-Separation Binary Systems in Synoptic Sky Surveys}

\author{Emil Terziev\altaffilmark{1}, Nicholas M. Law\altaffilmark{2}, Iair Arcavi\altaffilmark{3}, Christoph Baranec\altaffilmark{4}, Joshua S. Bloom\altaffilmark{5}, Khanh Bui\altaffilmark{4}, Mahesh P. Burse\altaffilmark{6}, Pravin Chorida\altaffilmark{6}, H.K. Das\altaffilmark{6}, Richard G. Dekany\altaffilmark{4}, Adam L. Kraus\altaffilmark{7}, S. R. Kulkarni\altaffilmark{4}, Peter Nugent\altaffilmark{8,5}, Eran O. Ofek\altaffilmark{9}, Sujit Punnadi\altaffilmark{6}, A. N. Ramaprakash\altaffilmark{6}, Reed Riddle\altaffilmark{4}, Shriharsh P. Tendulkar\altaffilmark{4}}

\altaffiltext{1}{Dunlap Institute for Astronomy and Astrophysics, University of Toronto, 50 St. George St., Toronto, Ontario, M5S 3H4, Canada; emil.terziev@utoronto.ca}
\altaffiltext{2}{Dunlap Fellow, Dunlap Institute for Astronomy and Astrophysics, University of Toronto, 50 St. George St., Toronto, Ontario, M5S 3H4, Canada}
\altaffiltext{3}{Department of Particle Physics and Astrophysics, The Weizmann Institute of Science, Rehovot 76100, Israel}
\altaffiltext{4}{Cahill Center for Astrophysics, California Institute of Technology, Pasadena, CA, 91125, USA}
\altaffiltext{5}{Department of Astronomy, University of California, Berkeley, CA 94720-3411, USA}
\altaffiltext{6}{Inter-University Centre for Astronomy \& Astrophysics, Ganeshkhind, Pune, 411007, India}
\altaffiltext{7}{Hubble Fellow, Institute for Astronomy, University of Hawaii, 2680 Woodlawn Drive, Honolulu, HI, 96822, USA}
\altaffiltext{8}{Computational Cosmology Center, Lawrence Berkeley National Laboratory, 1 Cyclotron Road, Berkeley, CA 94720, USA}S
\altaffiltext{9}{Benoziyo Center for Astrophysics, Weizmann Institute of Science, 76100 Rehovot, Israel.}

\begin{abstract}
The direct detection of binary systems in wide-field surveys is limited by the size of the stars' point-spread-functions (PSFs). A search for elongated objects can find closer companions, but is limited by the precision to which the PSF shape can be calibrated for individual stars. We have developed the BinaryFinder algorithm to search for close binaries by using precision measurements of PSF ellipticity across wide-field survey images. We show that the algorithm is capable of reliably detecting binary systems down to $\approx$1/5 of the seeing limit, and can directly measure the systems' position angles, separations and contrast ratios. To verify the algorithm's performance we evaluated 100,000 objects in Palomar Transient Factory (PTF) wide-field-survey data for signs of binarity, and then used the Robo-AO robotic laser adaptive optics system to verify the parameters of 44 high-confidence targets. We show that BinaryFinder correctly predicts the presence of close companions with a $<$5\% false-positive rate, measures the detected binaries' position angles within $2^\circ$ and separations within $25\%$, and weakly constrains their contrast ratios. When applied to the full PTF dataset, we estimate that BinaryFinder will discover and characterize $\sim$450,000 physically-associated binary systems with separations $<$$2$ arcseconds and magnitudes brighter than $\rm m_R=18$. New wide-field synoptic surveys with high sensitivity and sub-arcsecond angular resolution, such as LSST, will allow BinaryFinder to reliably detect millions of very faint binary systems with separations as small as 0.1 arcseconds.
\end{abstract}


\maketitle

\section{Introduction}
The development of ground-based high-angular-resolution imaging techniques such as adaptive optics (e.g. \citealt{Wizinowich2006, Herriot2000, Troy2000, Rousset2000, Moretti2009, Neichel2010, Hart2010}), Lucky Imaging (e.g. \citealt{Law2006_Lucky}) and aperture-masking (e.g. \citealt{Monnier2004, Lacour2011, Tuthill2000}) has been driven by the need to discover or characterize closely-separated objects, usually after initial target identification by a wide-field survey. Such follow-up is typically limited by oversubscription and telescope efficiency to at most several hundred targets per survey (for example, in the low-mass-star regime: \citealt{Siegler2005, Close2003, Law2010_wide, Janson2012}). New developments such as the high-efficiency robotic adaptive optics system Robo-AO \citep{Baranec2012} are increasing the possible sample sizes to thousands of targets. However, there remain many orders of magnitude more targets which cannot feasibly be searched one-by-one.

On the other hand, synoptic wide-field imaging covers millions of objects in each exposure, but with a seeing-limited resolution of 0.5--2 arcseconds. The Palomar Transient Factory \citep{Law2009, Rau2009}, Pan-Starrs \citep{Kaiser2002}, Skymapper \citep{Keller2007}, and ultimately LSST \citep{LSST2008, LSST2009} are, or will be, generating extremely large datasets covering billions of objects. 

In this paper we present BinaryFinder, a technique which is capable of efficiently searching wide-field synoptic survey data for close companions.  The technique uses algorithms developed for weak-lensing surveys \citep{Kaiser1995, Hoekstra1998, Hoekstra2005} to measure the ellipticity of individual target star images. We extend those techniques to obtain the probability of each imaged star having a close (0.1--2.0 arcseconds) companion and to determine the position angle, separation and contrast ratio of high-confidence binary systems.

Simple measurement of the ellipticity of stars in wide-field imaging data can reveal binaries of moderate separation (e.g. \citealt{Kraus2007}). However, imperfect optics and atmospheric effects lead to changing point spread functions (PSFs) across a wide field (PSF anisotropy), preventing the easy detection of close and/or high-contrast binaries. Our method uses hundreds of point sources in the field as calibration objects to measure PSF anisotropy, and multiple-epoch data to correct for the effects of changing seeing and to measure the separations and contrast ratios of the detected systems.

The paper is organized as follows. In Section \ref{sec:practical} we describe the PSF-ellipticity measurement and anisotropy correction algorithm, as well as the practical considerations of implementing the method with wide-field survey data. Section \ref{sec:testing} describes our method of generating artificial point sources and binary systems, applying simulated PSF anisotropies, accounting for instrumental effects, and matching the resulting ellipticity distributions to those seen on-sky. A verification of our method is presented in Section \ref{sec:roboao} where we use Robo-AO to confirm and characterize binaries discovered by our algorithm. In Section \ref{sec:characterize} we explore the use of multiple-epoch data to measure the position angle, separation and contrast ratios of detected binaries. We conclude in Section \ref{sec:concs} with a discussion of the false-positive rate of the algorithm, and the applications of the method.

\section{The BinaryFinder Algorithm}
\label{sec:practical}

The BinaryFinder algorithm can be divided into these steps:

\begin{enumerate}
\item{Measure the raw ellipticities and PSF anisotropy parameters of all stars in a wide-field seeing-limited image (Sections \ref{sec:algorithm} \& \ref{sec:obj_det}).}
\item{Divide the image into smaller subsections and bin the stars within each subsection into groups of similar flux (Section \ref{sec:image_part}). For each of these groups, derive a polynomial fit which describes how the PSFs vary as a function of position on the CCD, and use the fit to correct the raw ellipticity measurements to obtain anisotropy-corrected ellipticities (Section \ref{sec:aniso_fit}).}
\item{Repeat this process for multiple exposures of the same field, providing several ellipticity measurements for each object in the field; consolidate the individual ellipticity measurements of each object by finding the seeing dependence of the ellipticity and calculate the ellipticity at a common reference seeing (Section \ref{sec:seeing}).}
\item{Evaluate each object for the presence of close companions (Sections \ref{sec:suitable} \& \ref{sec:results}).}
\item{Use the changing ellipticity as a function of seeing to constrain the separations and contrast ratios of individual systems (detailed in Section \ref{sec:characterize}).}
\end{enumerate}

In the following sections we describe each component in detail, with a particular application to wide-field survey data taken by the Palomar Transient Factory. 

\subsection{Ellipticity and PSF anisotropy measurement}
\label{sec:algorithm}

\citet{Hoekstra2005} (hereafter HWU) describes a method to measure the ellipticity of a single object with corrections for PSF anisotropy, which we summarize here. HWU quantify the ellipticity of a PSF with ellipticity parameters \emph{e$_{1}$} and \emph{e$_{2}$}, which relate the second moments of flux as measured from the centroid of the object. 

\begin{equation}
e_{1} =\frac{I_{11} - I_{22}}{I_{11} + I_{22}}   ,   e_{2} =\frac{2I_{12} }{I_{11} + I_{22}}
\end{equation}

where

\begin{equation}
I_{11} = \sum\limits_{x,y}{f(x,y)x^{2}W} 
\end{equation}
\begin{equation}
I_{22} =  \sum\limits_{x,y}{f(x,y)y^{2}W} 
\end{equation}
\begin{equation}
I_{12} =  \sum\limits_{x,y}{f(x,y)xyW} 
\end{equation}

Here, \emph{x} and \emph{y} are the x and y pixel numbers as measured in a coordinate system whose origin is at the centroid of the object, \emph{f(x,y)} is the source flux falling on the pixel located at the point \emph{(x,y)} in that same coordinate system , and \emph{W} is a weight function which accounts for the increase in Poisson noise with decreasing photon count away from the centroid. 

The ellipticity of the PSF of an unresolved binary or an extended object is a combined measure of the true ellipticity caused by the astrophysical configuration of the object, and the ellipticity induced by directional smearing of starlight by the atmosphere and telescope optics. Thus, any raw ellipticity parameters need to be corrected for the induced ellipticity in order to obtain astrophysically significant measures of the true ellipticity. HWU do this using the smear polarizability tensor \emph{P} which quantifies the changes in PSF ellipticity resulting from perturbations in the PSF anisotropy.  

\begin{equation}
P_{ij}=X_{ij}-e_{i}e_{j}^{sm}
\end{equation}

where 
\begin{equation}
X_{11}=\frac{1}{I_{11} + I_{22}}  \sum\limits_{x,y}{f(x,y)(W+2W'(x^{2}\!\!+\!\!y^{2})+W''(x^{2}\!\!-\!\!y^{2}))} 
\end{equation}

\begin{equation}
X_{22}=\frac{1}{I_{11} + I_{22}} \sum\limits_{x,y}{f(x,y)(W+2W'(x^{2}\!\!+\!\!y^{2})+4W''x^{2}y^{2})}      
\end{equation}

\begin{equation}
e_{1}^{sm}=\frac{1}{I_{11} + I_{22}} \sum\limits_{x,y}{f(x,y)(x^{2}-y^{2})(2W'+W''(x^{2}+y^{2}))}       
\end{equation}

\begin{equation}
e_{2}^{sm}=\frac{1}{I_{11} + I_{22}} \sum\limits_{x,y}{f(x,y)(2xy)(2W'+W''(x^{2}+y^{2}))}    
\end{equation}

Here, the differentiation of the weight function \emph{W} is taken to be with respect to \emph{x$^{2}$+y$^{2}$}. 

The smear polarizability tensor of a source allows HWU to measure the PSF anisotropy parameters \emph{p$_{1}$} and \emph{p$_{2}$} which quantify the anisotropy in the image. 
\begin{equation}
p_{i}=\frac{e_{i}}{P_{ii}}
\end{equation}

To correct the ellipticity of a star in an image, it is necessary to examine how the anisotropy parameters vary across the CCD in the vicinity of the star. HWU measure the anisotropy parameters of point sources near the star of interest, and create a polynomial fit of \emph{p$_{1}$} and \emph{p$_{2}$} as a function of position on the CCD. They then correct the observed ellipticity of the star via the relationship

\begin{equation}
e_{i}^{cor}=e_{i}^{obs}-P_{ii}p_{i}-\alpha (p_{1}^{2}+p_{2}^{2})^{0.5}
\end{equation}

where $\alpha$
 is a constant proportional to the magnitude of the anisotropy at the point on the CCD where the star is located. This provides a final, anisotropy-corrected measure of the PSF ellipticity. 

HWU measure the ellipticity of the object in many separate observations. The observations are taken at varying seeings, thus they determine the ellipticity at a reference FWHM by deriving a fit of ellipticity as a function of FWHM and evaluating the fit at the chosen reference radius. 

\subsection{Object detection and measurement of raw PSF parameters}
\label{sec:obj_det}
In order to measure the necessary parameters of all the stars in our images (using Equations 1-10), we require centroid coordinates and Full-Width at Half-Max (FWHM) measurements of the PSF of each source. We obtain a catalog of objects (these include stars, blended binary systems, foreground/background star blends, and galaxies) for each image using SExtractor \citep{Bertin1996}. SExtractor provides the sky coordinates (RA and DEC) and pixel coordinates (x and y) of the centroid of each object, the FWHM assuming a Gaussian core, the flux, and the SExtractor error flag. We remove all objects with non-zero SExtractor flags to avoid saturation, bad pixels, and the wider-separation binaries which SExtractor can directly detect and deblend.

Using the catalog of objects obtained from SExtractor, we measure the raw \emph{e$_{1}$} and \emph{e$_{2}$} ellipticity parameters and the PSF anisotropy parameters \emph{p$_{1}$} and \emph{p$_{2}$} of every object in an image. The weight function we employ to account for changing SNR across the PSF is a Gaussian with dispersion equal to the FWHM calculated by SExtractor, centered at the SExtractor-derived centroid coordinates. We define an object aperture with a radius equal to twice the FWHM, and a sky-subtraction aperture with radii between 2.5$\times$ and 3.5$\times$ the FWHM.

\subsection{Image partitioning}
\label{sec:image_part}
\begin{figure}
\includegraphics[width=\columnwidth]{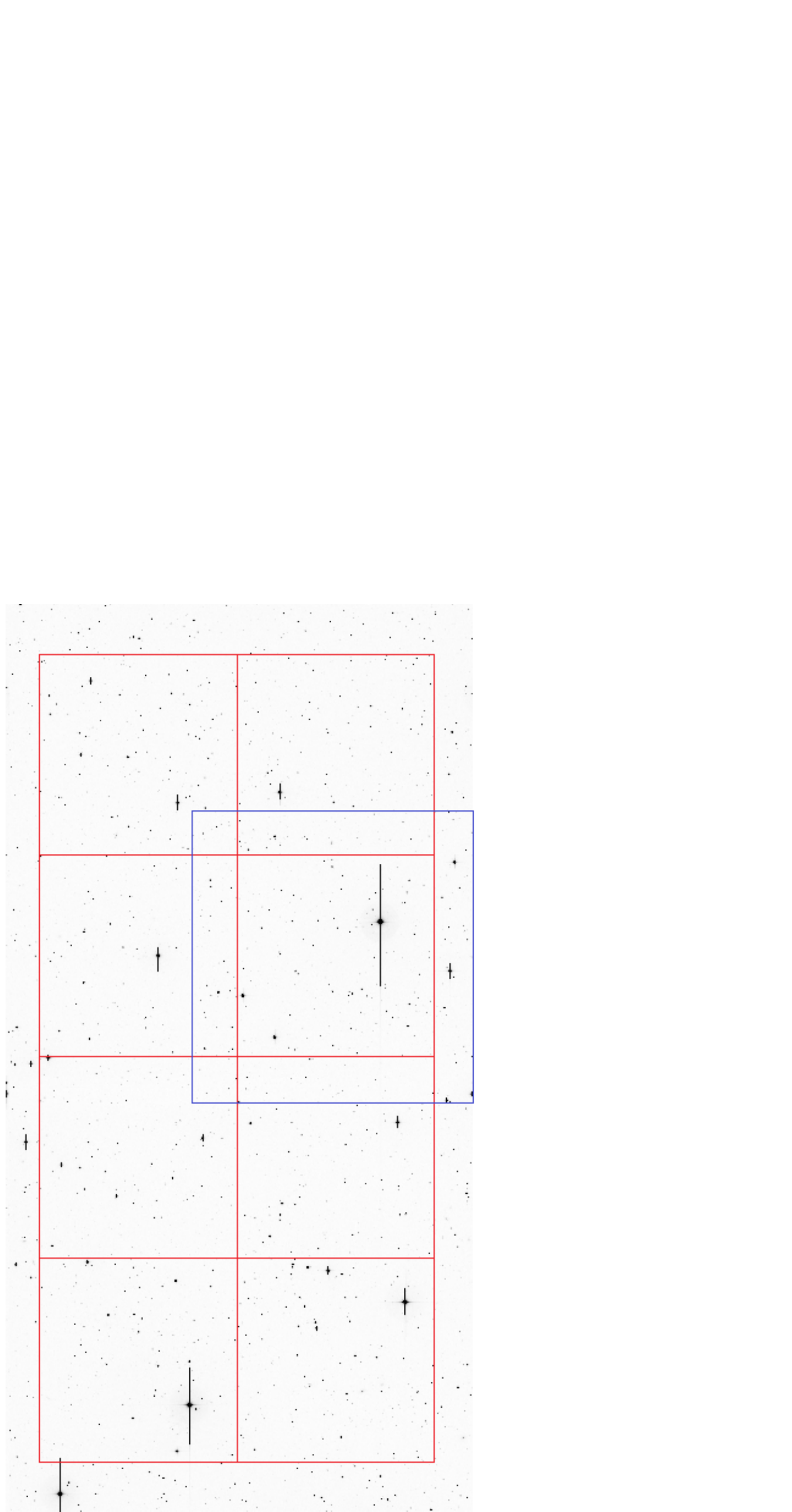}
\caption{Partitioning of image (part of PTF dataset). The images are 2048 by 4096 pixels with a 1.01'' pixel size. The red grid divides the image into 8 smaller subsections about 800 by 800 pixels in size, each of which will be corrected for anisotropy with a separate polynomial fit. The larger (1200x1200) blue square which encloses one of the partitions of the red grid shows the region from which the fit will be derived for that particular partition. }
%%\label{Figure 1}
\label{fig:figure1}
\end{figure}

The anisotropy profile of images can vary significantly in wide-field images due to geometric distortions. In PTF data, we found it necessary to partition the images into subsections, and derive separate fits for each partition (See Figure \ref{fig:figure1}).  It is important to make the regions as close to square in shape as possible, as an accurate fit requires a comparable span in each direction. Additionally, the subset of sources used to create the fits should always extend further than the boundaries of the region being corrected (as shown by the blue square in Figure \ref{fig:figure1}), to ensure that the ellipticities of all sources are corrected based on a fit obtained from sources surrounding them on all sides.

\subsection{PSF anisotropy fit}
\label{sec:aniso_fit}
We next derive a polynomial fit of the anisotropy parameters for each group to determine how the PSF anisotropy varies across the CCD. We perform a 3$\sigma$ clip on the values of \emph{e$_{1}$}, \emph{e$_{2}$}, \emph{p$_{1}$} and \emph{p$_{1}$} to remove severe outliers. The points that remain may not all be point sources (some are blends, galaxies, etc). In contrast to instrumentally-induced ellipticity, this will not cause any directional bias in the fit, as we expect astrophysically-extended objects to appear with equal probabilities in all orientations. 

Our anisotropy fit is of the form
\begin{equation}
p_{1}^{fit}=c_{1}x+c_{2}x^{2}+c_{3}y+c_{4}y^{2}+c_{5}xy+c_{6}+c_{7}f_{tot}+c_{8}f_{tot}^{2}
\end{equation}
where \emph{c$_{i}$} are the constant coefficients to be found which minimize the residuals of the measured \emph{p$_{1}$} parameters of the objects being fitted and \emph{x} and \emph{y} are the CCD pixel coordinates where the objects are located. 

We extended the \citet{Hoekstra2005} model to include terms which are proportional to the flux of the objects (\emph{f$_{tot}$}). While applying the algorithm to the PTF images, we noticed that the ellipticities of high-flux objects were not corrected accurately to zero using a fit dependent only on x and y object coordinates. Upon closer examination of the images, we realized that for high flux (yet unsaturated) objects, slightly-less-than 100\% Charge Transfer Efficiency (CTE) leads to a smearing in the direction of readout. This causes an additional elongation dependent on the flux of the object, requiring flux terms in the anisotropy fit.

A second-order polynomial fit appears to be sufficient for PTF images; first degree polynomial fits were insufficient to accurately quantify the polarization variation, and we found that third degree and higher fits do not increase the accuracy of the corrected parameters compared to the second order fit.

A similar fit is also derived for the  \emph{p$_{2}$} parameter, although the flux-dependent terms are not necessary as readout-induced elongation exactly parallel to the pixel grid (whether it is the x or y axis) does not affect the  \emph{e$_{2}$} parameter, by definition (see equations 1 through 4). 

Once we have derived these fits we evaluate them at the coordinates of the objects to be corrected and obtain the corrected ellipticities according to Equation 12. For the purposes of evaluating the ellipticity of an object compared to others in the field, we use \emph{e}, the $\sqrt{e_1^2 + e_2^2}$ combination of the ellipticities.

We find that the typical instrument-induced ellipticity has a magnitude between 0.01 and 0.1, roughly the same amount as our final measured ellipticities for detected binary systems. 

\subsection{Seeing correction}
\label{sec:seeing}
\begin{figure}
\includegraphics[width=\columnwidth]{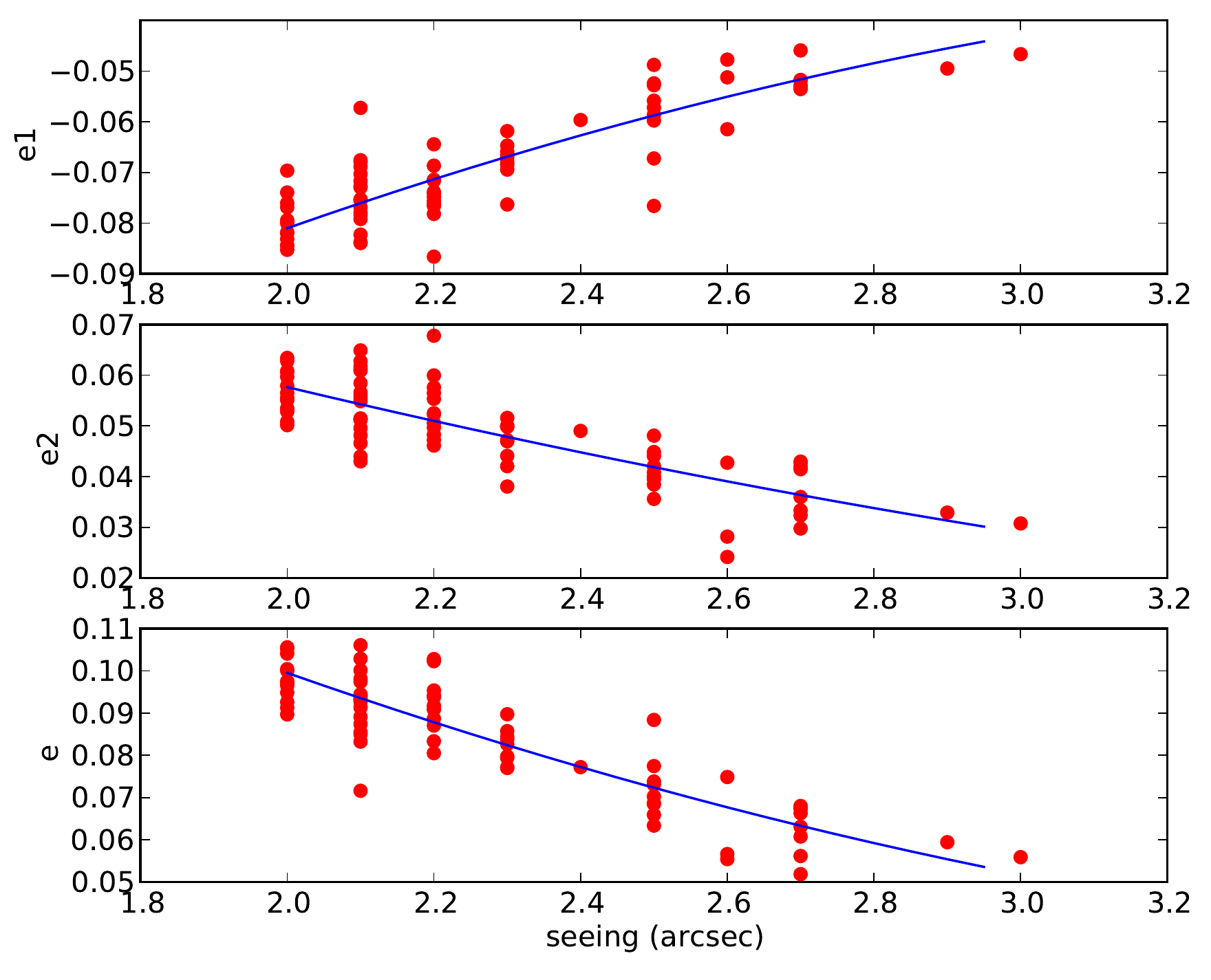}
\caption{Seeing fit of \emph{e$_{1}$}, \emph{e$_{2}$} and \emph{e} of an elongated PTF object. The points represent the various ellipticity measurements obtained from each unique image of the PTF field the object is in. The curve shows the polynomial fit of ellipticity as a function of the FWHM. The fits of all objects are evaluated at a common reference FWHM to make ellipticities comparable across fields.}
\label{fig:seeingdepe1ande2a}
\end{figure}

When we measure the corrected ellipticities of an object across multiple images, we measure many values for \emph{e$_{1}$} and \emph{e$_{2}$} at a variety of FWHMs. However, ellipticity is not constant with seeing; close objects induce smaller ellipticities when the PSFs are large. To make ellipticity measurements comparable between fields with different seeing distributions, we create a fit of object ellipticities as a function of FWHM, and evaluate this fit at a reference seeing value (Figure \ref{fig:seeingdepe1ande2a}). 

\subsection{Selecting suitable fields and objects}
\label{sec:suitable}
Undersampled images with FWHMs close to the pixel scale of the CCD cannot accurately measure ellipticities because the ellipticity measurement is then very sensitive to the location of the object relative to the pixel grid. We found that pixel effects become negligible for roughly Nyquist-sampled images with at least two pixels across the star PSFs, and we therefore use only those well-sampled PTF images. We require at least 30 separate observations of a target in order for us to obtain an accurate fit of ellipticity with seeing, and we also require several hundred sufficiently bright sources within our image for PSF measurement (the last requirement is satisfied by essentially all PTF images). For R-band PTF images, our simulations and on-sky measurements showed that the faintest useful sources have $\rm m_R\sim18$.

\subsection{Results for a typical wide-field-survey image}
\label{sec:results}
\begin{figure}
\includegraphics[width=\columnwidth]{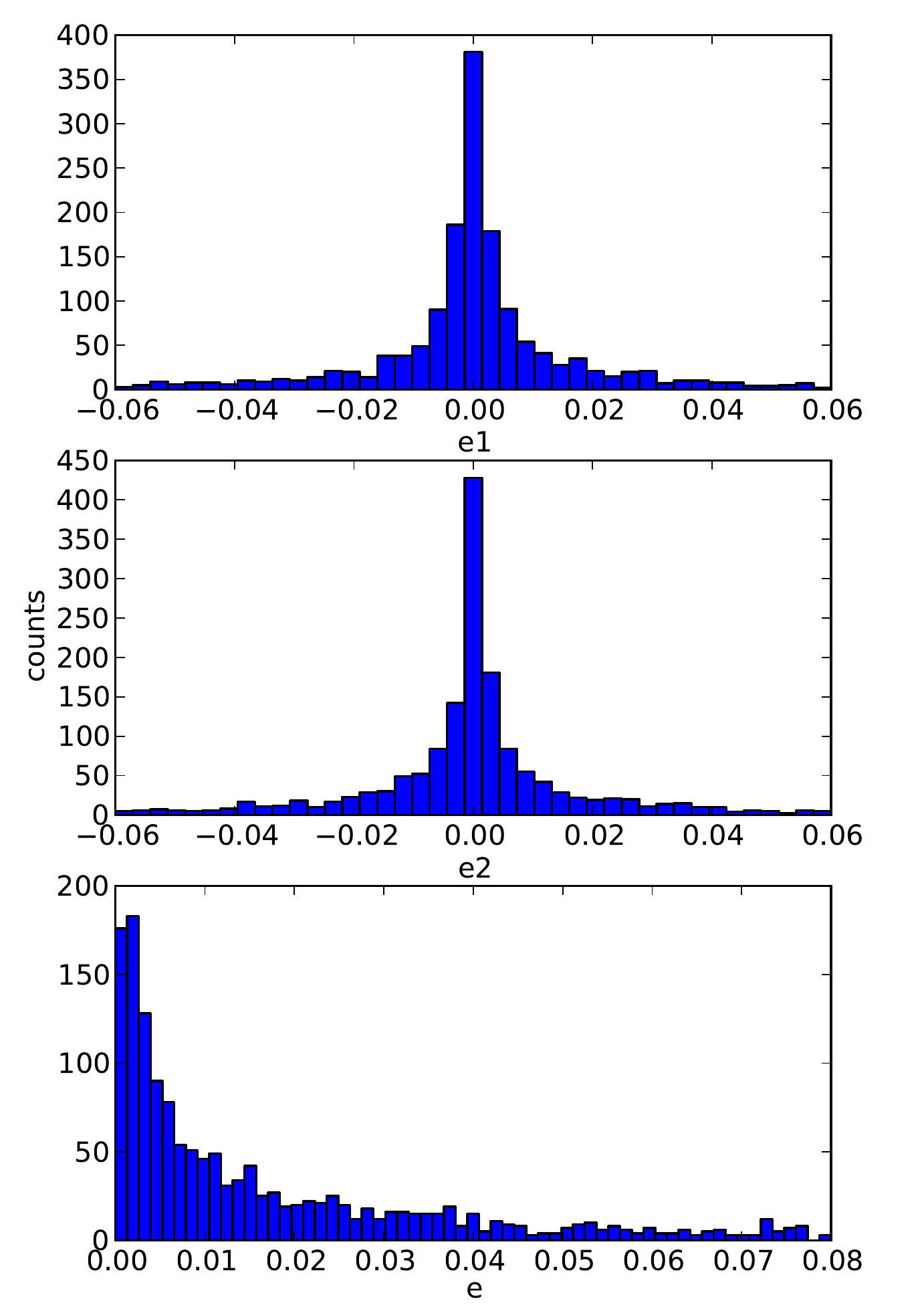}
\caption{Distributions of  \emph{e$_{1}$}, \emph{e$_{2}$} and \emph{e} of objects in a PTF field. The ellipticity parameters have all been evaluated at a reference FWHM of 2.5 arcseconds}
\label{fig:onefield}
\end{figure} 
\begin{figure}
\includegraphics[width=\columnwidth]{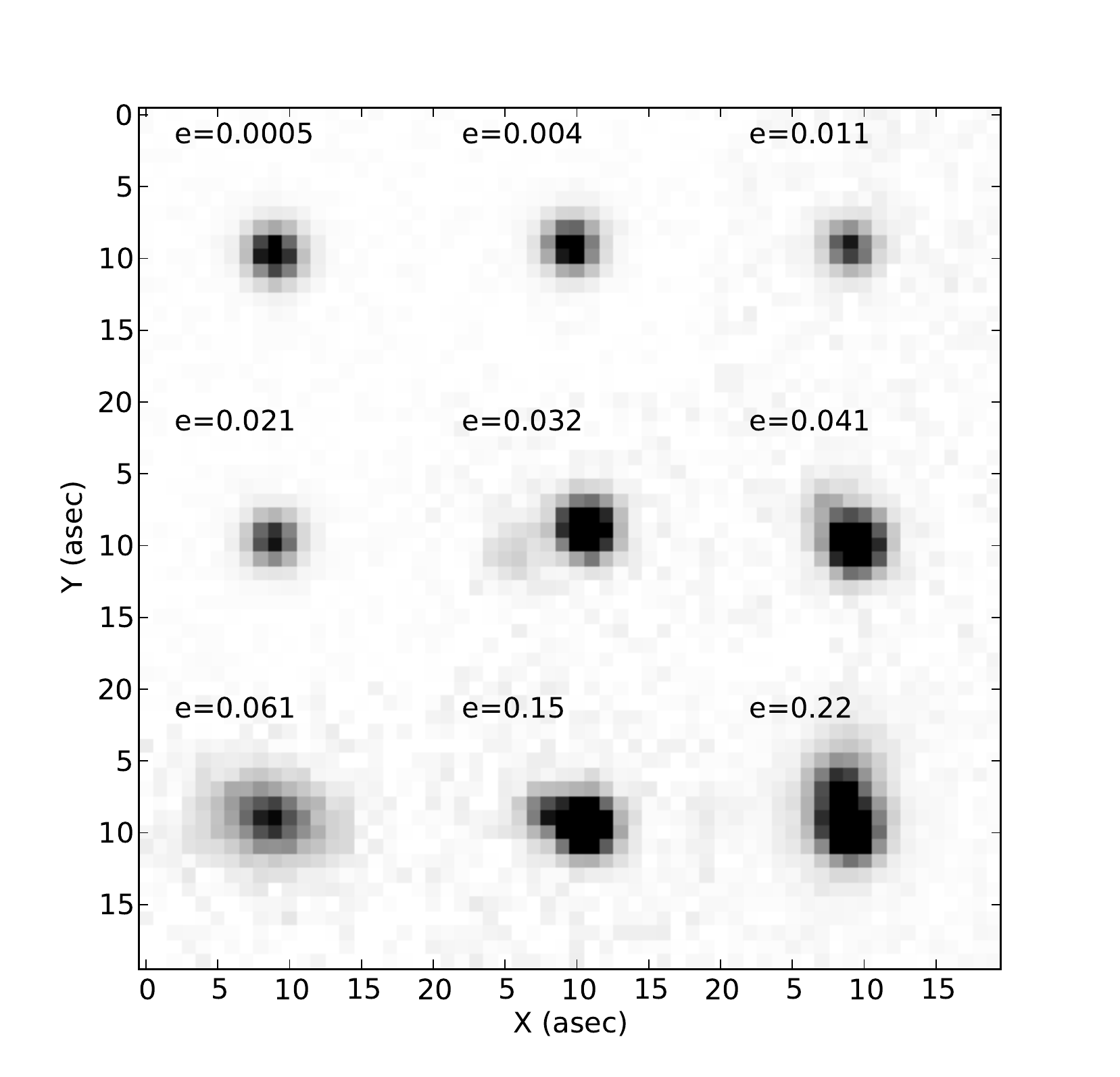}
\caption{A sample of PTF objects ordered by increasing ellipticity. Each image is 20 by 20 arcseconds in size and has a FWHM of $\sim$2.4 arcseconds.}
\label{fig:objectexamples}
\end{figure} 

Figure \ref{fig:onefield} provides an example of the distribution of \emph{e$_{1}$}, \emph{e$_{2}$} and \emph{e} (all measured at a common reference FWHM) of the objects appearing in a single chip of a PTF field (1/12 of a PTF field). The distributions have two components -- the bulk of the objects which are closely centered around zero ellipticity, and the extended wings. The first component represents true point sources or blends of objects too close together for our algorithm to differentiate from point sources. The thickness of this component of the distribution is due to measurement errors in the ellipticities. The wings of the distribution represent either blends of multiple point sources, or other extended objects such as galaxies. Figure \ref{fig:objectexamples} shows a sample of PTF objects from the same chip, ordered by increasing ellipticity.

\section{Simulated Performance}
\label{sec:testing}

\begin{figure}
\includegraphics[width=1.1\columnwidth]{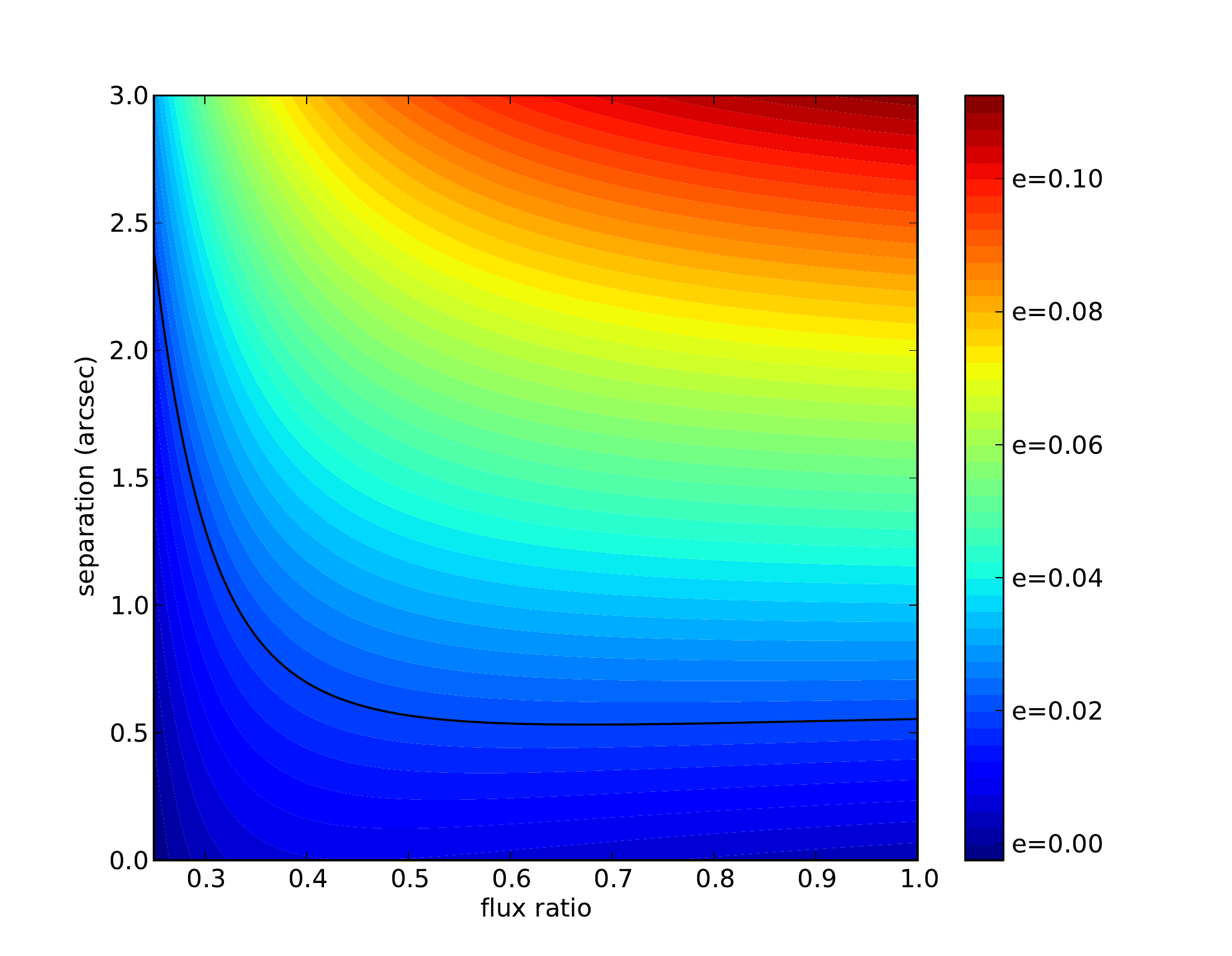}
\caption{Separation and contrast dependence of the ellipticity of the PSF of blended binaries (artificial images). The black curve represents the binary detection criterion ($e$=0.02).}
%%\label{Figure 4}
\label{fig:crit}
\end{figure}
We tested the sensitivity of BinaryFinder using simulated wide-field-survey data. For each simulated image we generated 2000 point sources, along with artificial binary systems of varying separations and brightness ratios. We modelled the PSF of each object using a Moffat function. Then, we convolved all the object PSFs with anisotropic kernels to simulate the instrumental PSF anisotropy. In order to make this anisotropy dependent on the location within the CCD, we placed four reference anisotropy kernels at the four corners of our CCD. We convolved each object with a kernel which was the weighted average of the four corner kernels, with the weights proportional to the inverse of the distance from the object centroid to that corner of the CCD. We then applied BinaryFinder, using the artificial point sources to derive the anisotropy fits and correct the raw ellipticities.  The changes in the ellipticity measurement with the brightness ratio and separation of our artificial binaries is shown in Figure \ref{fig:crit}. As expected, the anisotropy-corrected ellipticity increases for binaries of greater separation and greater brightness ratios, and vice versa. 

\begin{figure}
\includegraphics[width=\columnwidth]{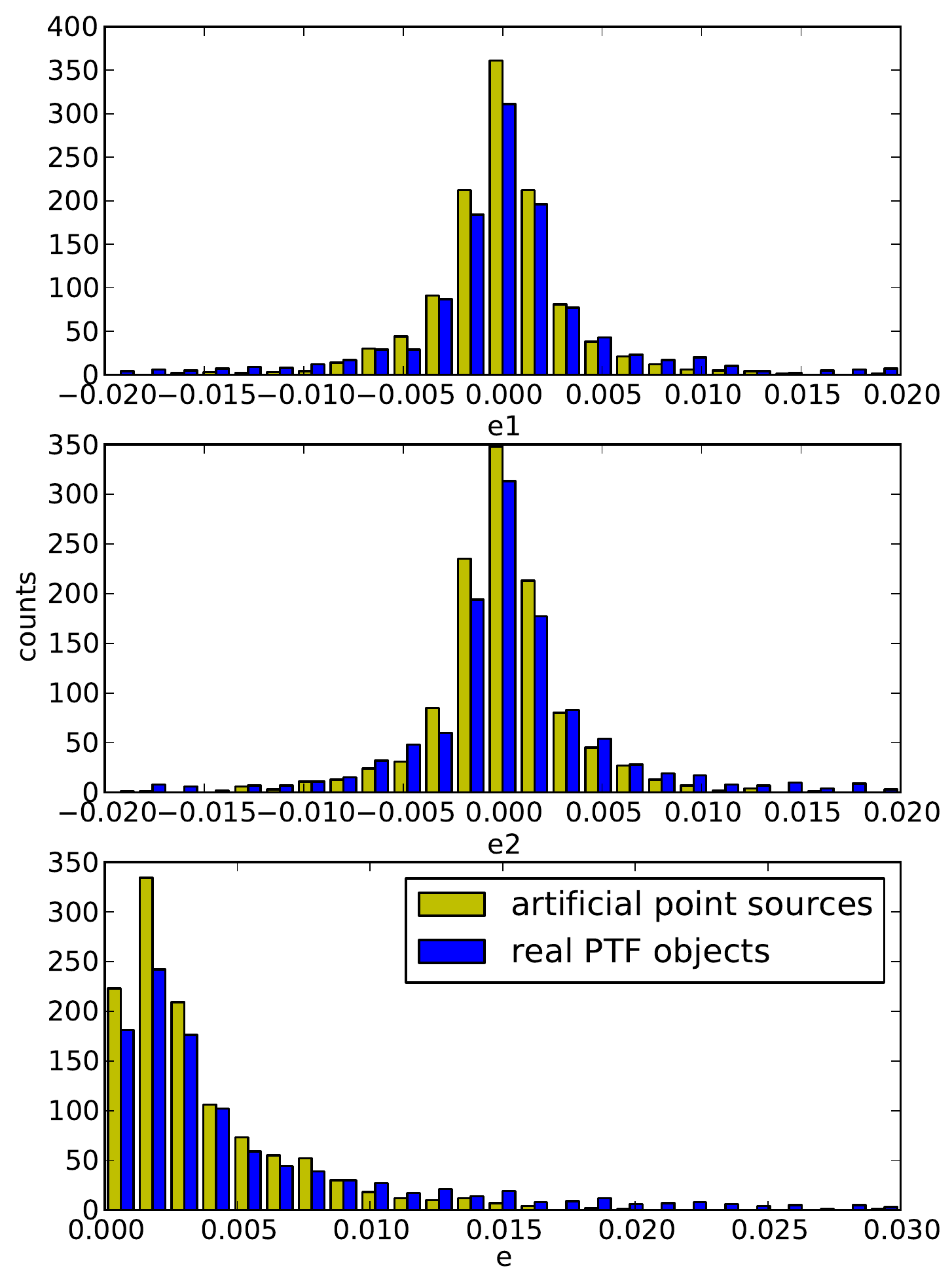}
\caption{Ellipticity distributions of 1200 artificial point sources and 1200 randomly chosen PTF objects (top - $e_1$, middle - $e_2$ , bottom - $e$). As can be seen in the histograms, the frequency of the artificial point sources goes to zero for magnitudes of e1 or e2 greater than $\approx$0.015 or for $e$ greater than $\approx$0.02. In contrast, the real PTF object distributions have extended tails beyond those ellipticity values.}
%\label{Figure 5}
\label{fig:psspread3}
\end{figure}

We also utilized artificially generated point sources to determine an ellipticity boundary point between likely point sources and secure binary detections. We chose 1200 PTF objects at random, and created 1200 similar artificial point sources using the method outlined above. For the artificial sources, we simulated the observation conditions and data reduction of the real images, including variations in cloud cover and background brightness which change the signal to noise ratio from image to image. We also introduced a per-pixel flat field noise in the simulated images at the 1.4\% level. We plotted the ellipticities of the artificial point sources and the real objects (see Figure \ref{fig:psspread3}) to compare the two distributions. The simulations accurately reproduce the PTF-measured distribution (within sampling noise) in the area corresponding to point sources. However, the PTF objects have a much larger tail in ellipticity, corresponding to astrophysically-extended objects. We found that the frequency of simulated point sources falls to zero for $e$ above 0.02 in our simulations, and we adopt that as the boundary point between the point sources and the wide separation binaries. Note that the ellipticity values of the boundary point are pixel scale dependent and thus have to be calculated independently for new datasets.

\section{Verifying BinaryFinder Discoveries with Robo-AO adaptive optics imaging}
\label{sec:roboao}
\begin{figure}
\includegraphics[width=\columnwidth]{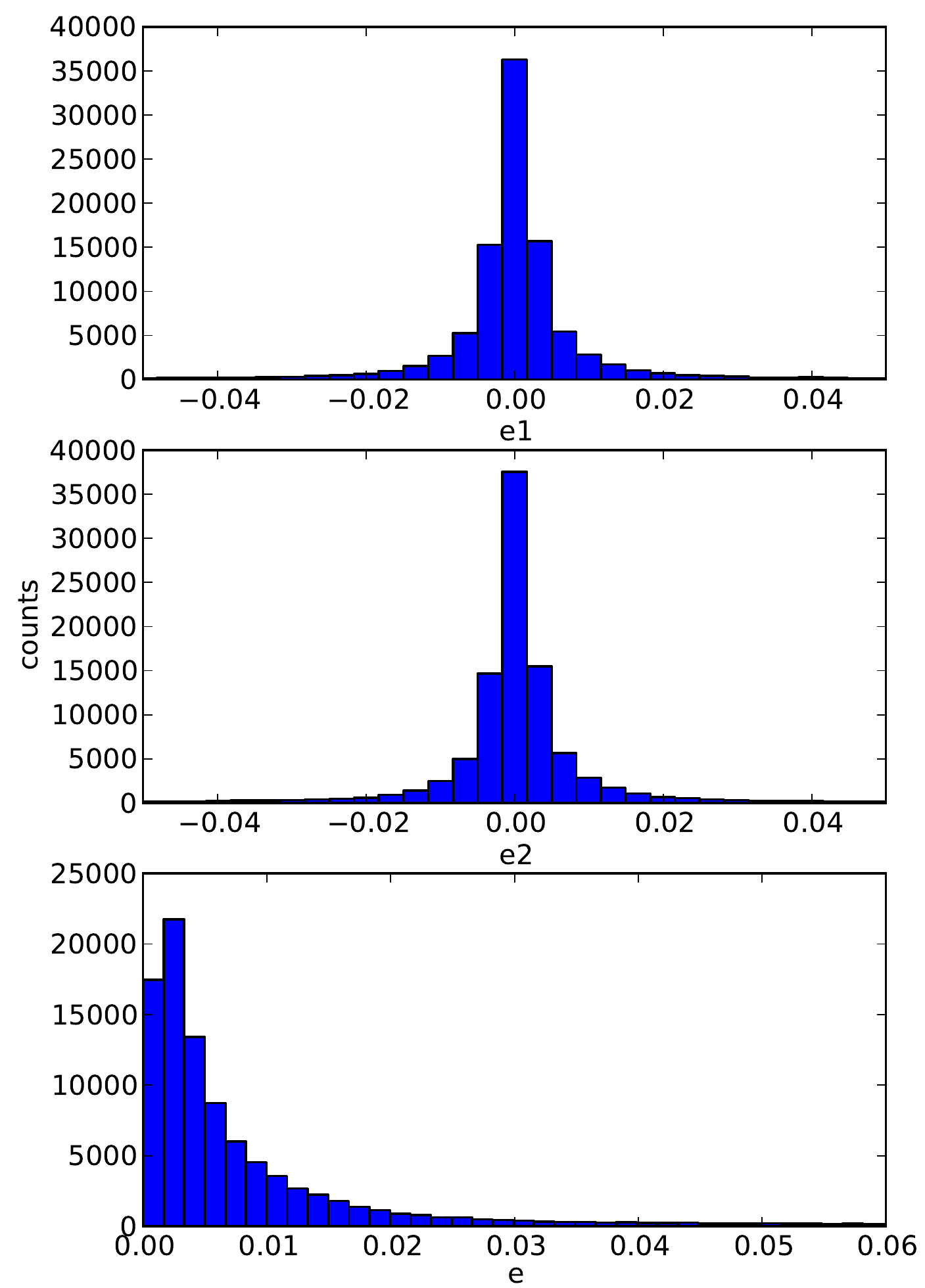}
\caption{Ellipticity distributions of the 100,000 PTF targets in the Robo-AO binary discovery target area.}
%\label{Figure 6}
\label{fig:ptfdist}
\end{figure}

To verify the performance of the BinaryFinder algorithm we used the Robo-AO robotic laser adaptive optics system to obtain 0.1-arcsecond resolution images of a large sample of possible binaries. To generate the target list we measured the astrophysical ellipticities of approximately 100,000 objects in PTF data covering a $\approx$70 square degree area (see Figure \ref{fig:ptfdist} for the distribution of ellipticities in this region). To limit our ellipticity search to stellar binaries, we measured the proper motion for each object in our sample from the USNO-B1, 2MASS and (where available) SDSS measured positions (\citealt{Monet2003, Skrutskie2006, York2000}). We required at least a $2.5\sigma$ proper motion detection for selection for observation followup, with target brightness limits of $\rm14<m_{R}<16.0$, set by the PTF saturation limit and Robo-AO requirements. We selected 44 sources at a wide range of ellipticities from the resulting target list for follow-up observations with Robo-AO.

\subsection{Robo-AO} 
Robo-AO is a visible and near-infrared laser guide star adaptive optics system specifically engineered for 1--3 m class telescopes \citep{Baranec2012}. The Robo-AO system located at the Palomar 60-inch telescope comprises an ultraviolet Rayleigh laser guide star, an integrated adaptive optics and science camera system, and a robotic control system. The system currently incorporates both an electron-multiplying CCD and an InGaAs infrared array camera for imaging. The robotic system is designed for high-efficiency observing, allowing large numbers of targets to be imaged rapidly. The diffraction-limited visible-light capability of Robo-AO allowed us to make a similar-wavelength comparison between the PTF-measured ellipticities and the observed binary contrast ratios.

\begin{figure*}
\begin{center}

\includegraphics[height=1.0\textheight]{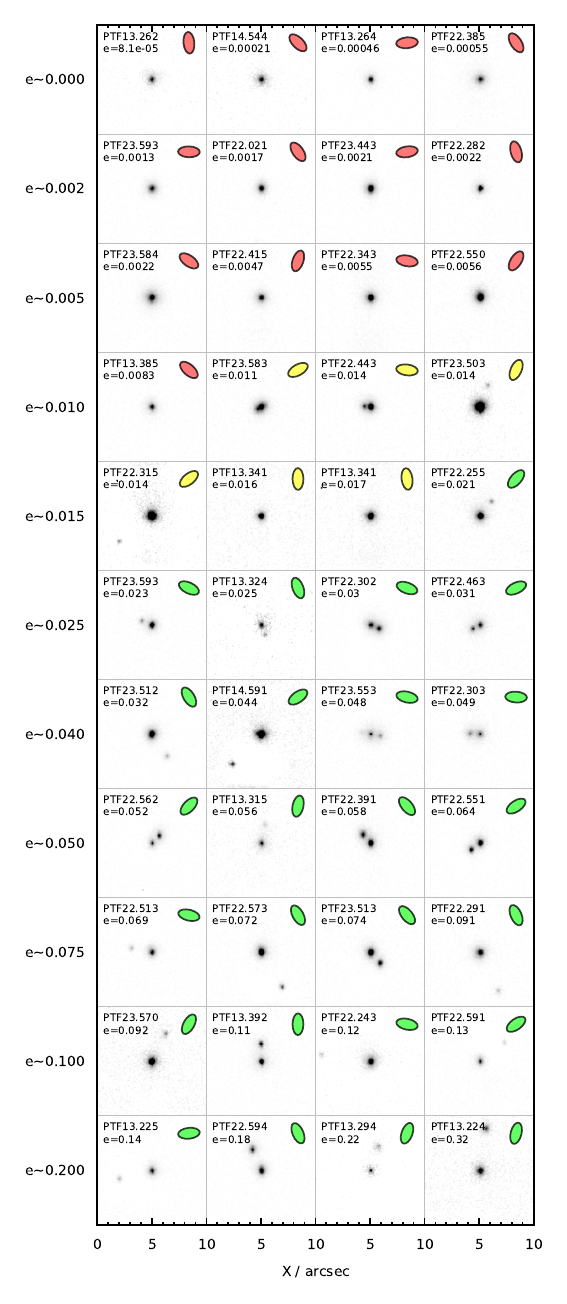}
\caption{Robo-AO adaptive-optics images of the binarity test targets. Each image is 10 by 10 arcseconds in size; East is towards the left and the images are rotated by 23.5$^{\circ}$ counter-clockwise with respect to a North-East axis. The images are shown with linear scaling with levels selected to best display the binarity (or lack thereof) of each of the targets. The ellipses indicate the predicted binary orientation based on the relative magnitudes of the \emph{e$_{1}$} and \emph{e$_{2}$} parameters; the color indicates confidence level of the predicted companion (red: no companion; yellow: possible companion; green: very likely companion). The order of targets is the same as that of Table 1.}

\label{robo_ao}
\end{center}
\end{figure*}

\begin{deluxetable*}{lllrrrlll}
\tabletypesize{\footnotesize}
%\rotate
\tablecaption{Robo-AO test targets, ordered by increasing PTF-measured ellipticity}
\tablewidth{0pt}
\tablehead{
\colhead{Target Name} & \colhead{\shortstack{RA\\ (J2000)}}	 & \colhead{\shortstack{DEC\\(J2000)}} & \colhead{\emph{e$_{1}$}} & \colhead{\shortstack{\emph{e$_{2}$}}} & \colhead{\shortstack{\emph{e}}} & \colhead{\shortstack{separation\\(arcsec)}} & \colhead{\shortstack{r-band contrast\\(flux ratio)}} 
}
\startdata

PTF13.262 &   13:26:29.9 &   +14:43:49 & -6.8e-05 & -4.3e-05 &  8.1e-05 & \nodata&        \nodata   \\
PTF14.544 &   14:54:47.0 &    +36:27:00 & -0.00014 &  0.00016 &  0.00021 & \nodata&        \nodata   \\
PTF13.264 &   13:26:44.1 &    +14:32:09 & 0.00038 &  0.00025 &  0.00046 & \nodata&        \nodata   \\
PTF22.385 &  22:38:56.1 &  +18:44:41 &  -0.0005 &  0.00021 &  0.00055 & \nodata&        \nodata   \\
PTF23.593 &  23:59:37.6 &  +27:48:55 &  0.00091 &  0.00093 &   0.0013 & \nodata&        \nodata   \\
PTF22.021 &  22:02:15.3 &  +21:27:53 &  -0.0015 &  0.00078 &   0.0017 & \nodata&        \nodata   \\
PTF23.443 &  23:44:35.1 &  +28:00:39 &   0.0019 &  0.00092 &   0.0021 & \nodata&        \nodata   \\
PTF22.282 &  22:28:29.7 &  +18:07:36 &  -0.0021 & -0.00046 &   0.0022 & \nodata&        \nodata   \\
PTF23.584 &  23:58:40.8 &  +29:06:46 & -0.00086 &    0.002 &   0.0022 & \nodata&        \nodata   \\
PTF22.415 &  22:41:53.8 &  +18:10:51 & -0.00076 &  -0.0046 &   0.0047 & \nodata&        \nodata   \\
PTF22.343 &  22:34:32.6 &   +18:22:01 &   0.0026 &   0.0048 &   0.0055 & \nodata&        \nodata   \\
PTF22.550 &  22:55:04.1 &  +22:36:37 &   0.0017 &  -0.0053 &   0.0056 & \nodata&        \nodata   \\
PTF13.385 &   13:38:57.8 &   +41:18:25 &  -0.0049 &   0.0067 &   0.0083 & \nodata&        \nodata   \\
PTF23.583 &  23:58:39.2 &  +27:05:38 &     0.01 &  -0.0027 &    0.011 & $0.40\pm0.04$ & $0.45\pm0.07$  \\
PTF22.443 &  22:44:30.6 &  +18:34:25 &   0.0074 &    0.012 &    0.014 & $0.58\pm0.04$ & $0.28\pm0.03$  \\
PTF23.503 &    23:50:34.0 &  +29:07:39 &  0.00022 &   -0.014 &    0.014 & $2.01\pm0.04$ & $0.04\pm0.04$  \\
PTF22.315 &  22:31:58.8 &  +18:44:45 &    0.012 &  -0.0079 &    0.014 & $3.78\pm0.04$ & $0.07\pm0.06$  \\
PTF13.341 &   13:34:18.4 &   +13:08:51 &   -0.012 &    -0.01 &    0.016 & $0.20\pm0.04$* & $0.30\pm0.01$*  \\
PTF13.341 &   13:34:15.6 &   +13:10:38 &   -0.015 &  -0.0078 &    0.017 & \nodata&        \nodata   \\
PTF22.255 &  22:25:54.1 &  +19:57:30 &    0.012 &   -0.017 &    0.021 & $1.60\pm0.04$ & $0.11\pm0.02$  \\
PTF23.593 &  23:59:39.5 &  +28:43:27 &  -0.0018 &    0.023 &    0.023 & $0.98\pm0.04$ & $0.17\pm0.02$  \\
PTF13.324 &   13:32:46.8 &    +14:17:04 &   -0.025 & -0.00032 &    0.025 & $0.96\pm0.04$ & $0.26\pm0.07$  \\
PTF22.302 &  22:30:25.1 &  +18:22:13 &   0.0041 &     0.03 &     0.030 & $0.79\pm0.04$ & $0.89\pm0.01$  \\
PTF22.463 &   22:46:36.0 &  +18:59:59 &    0.031 &  -0.0037 &    0.031 & $0.73\pm0.04$ & $0.52\pm0.01$  \\
PTF23.512 &  23:51:23.3 &  +28:42:53 &    -0.03 &    0.011 &    0.032 & $2.42\pm0.04$ & $0.12\pm0.02$  \\
PTF14.591 &   14:59:17.3 &   +36:23:51 &    0.039 &    -0.02 &    0.044 & $3.77\pm0.04$ & $0.18\pm0.02$  \\
PTF23.553 &  23:55:35.6 &   +29:01:03 &    0.018 &    0.044 &    0.048 & $0.84\pm0.04$ & $0.43\pm0.02$  \\
PTF22.303 &  22:30:38.4 &   +19:53:04 &    0.032 &    0.038 &    0.049 & $0.91\pm0.04$ & $0.50\pm0.02$  \\
PTF22.562 &  22:56:24.9 &  +23:15:57 &    0.034 &   -0.039 &    0.052 & $0.94\pm0.04$ & $0.91\pm0.01$  \\
PTF13.315 &   13:31:53.6 &   +15:16:00 &   -0.022 &   -0.052 &    0.056 & $1.64\pm0.04$ & $0.17\pm0.06$  \\
PTF22.391 &  22:39:14.8 &  +19:17:56 &   -0.047 &    0.035 &    0.058 & $1.07\pm0.04$ & $0.64\pm0.01$  \\
PTF22.551 &  22:55:16.4 &  +22:30:49 &    0.058 &   -0.027 &    0.064 & $1.01\pm0.04$ & $0.55\pm0.01$  \\
PTF22.513 &  22:51:38.8 &  +23:02:19 &    0.024 &    0.064 &    0.069 & $1.92\pm0.04$ & $0.16\pm0.02$  \\
PTF22.573 &  22:57:37.7 &  +22:58:29 &   -0.069 &    0.018 &    0.072 & $3.67\pm0.04$ & $0.18\pm0.01$  \\
PTF23.513 &  23:51:30.5 &  +28:26:14 &    -0.06 &    0.043 &    0.074 & $1.25\pm0.04$ & $0.48\pm0.01$  \\
PTF22.291 &    22:29:14.0 &    +19:22:24 &   -0.091 &   0.0062 &    0.091 & $3.81\pm0.04$ & $0.14\pm0.03$  \\
PTF23.570 &   23:57:08.2 &  +28:53:47 &    0.018 &    -0.09 &    0.092 & $2.75\pm0.04$ & $0.12\pm0.03$  \\
PTF13.392 &   13:39:21.8 &   +41:25:23 &   -0.078 &   -0.075 &     0.11 & $1.61\pm0.04$ & $0.57\pm0.01$  \\
PTF22.243 &  22:24:37.1 &  +19:55:23 &    0.046 &     0.11 &     0.12 & $4.49\pm0.04$ & $0.12\pm0.05$  \\
PTF22.591 &  22:59:16.7 &  +23:18:27 &     0.12 &   -0.061 &     0.13 & $2.74\pm0.04$ & $0.17\pm0.03$  \\
PTF13.225 &   13:22:59.8 &   +41:11:44 &     0.12 &    0.072 &     0.14 & $3.04\pm0.04$ & $0.17\pm0.03$  \\
PTF22.594 &  22:59:43.1 &   +23:53:08 &    -0.18 &    0.012 &     0.18 & $2.05\pm0.04$ & $0.62\pm0.01$  \\
PTF13.294 &   13:29:49.7 &   +41:06:13 &   -0.028 &    -0.22 &     0.22 & $2.29\pm0.04$ & $0.33\pm0.07$  \\
PTF13.224 &   13:22:41.6 &     +14:00:14 &     -0.1 &     -0.3 &     0.32 & $3.82\pm0.04$ & $0.43\pm0.02$  \\
\\
&&&\multicolumn{5}{r}{* -- companion not associated with PTF ellipticity}\\
\enddata

\end{deluxetable*}

\subsection{Binarity verification}
We obtained Robo-AO images of the 44 targets on the nights of July 16-18 and August 4-7 2012 (UT). The July targets, with RAs around 13 hours, were observed with 60-second total exposure times in the i-band filter; the August targets (RAs of 22-23 hours) used a long-pass filter with a 600\,nm cut-on to obtain increased signal compared to a bandpass filter.  We operated Robo-AO without tip-tilt correction, instead relying on post-facto shift-and-add processing of the individual frames and used the pipeline described by \citet{Law2009_ao, Law2012} to perform the image alignment and co-addition. The Robo-AO system provided diffraction-limited resolution ($\sim0.1$ arcsec at these visible wavelengths) for all observed targets, in 1-2 arcsecond seeing conditions; the entire target list was observed in a total of only $\sim$2.5 hours.

The high-resolution images of our test sample (Figure \ref{robo_ao}) confirm the expected variation of binarity with measured ellipticity. Targets with small ellipticities  (e$<$0.01) are all confirmed to be single stars. We found that five of the six targets with $0.01<e<0.02$  are binaries; all the systems in that range are high-contrast and/or $<$1 arcsecond separation.  Targets with $e>0.02$ are all confirmed to be binaries, with separations increasing and contrast ratios decreasing as the ellipticities increase. A single target, PTF13.341, is detected as a binary with a faint, close companion by Robo-AO, but with a different position angle and closer separation than that measured from PTF data; this is likely to be a close companion not measured by our PTF data and we do not include it in further analysis. 

The crossover point between single stars and high-confidence binaries occurs at the ellipticities predicted by our artificial point source simulations (Section \ref{sec:testing}), although the $>$80\% binarity fraction of the targets with ellipticities between 0.01 and 0.02, compared to the zero binarity fraction at lower ellipticities, suggests that a less conservative limit could be set for many science programs.

\section{Characterizing detected binaries: position angles, separations and contrast ratios}
\label{sec:characterize}
Beyond simple detection of binarity the measured ellipticity vector allows us to directly measure the binary's position angle. In a multiple-epoch survey, the variation in ellipticity with seeing enable us to constrain both the binary's separation and its contrast ratio.

\subsection{Position angles}
The detected Robo-AO binaries show that our ellipticity measurements can predict the orientation of the binary from the relative magnitudes of the \emph{e$_{1}$} and \emph{e$_{2}$} parameters (see the blue ellipses in Figure \ref{robo_ao}). The \emph{e$_{1}$} parameter varies as cos(2$\theta$) and the \emph{e$_{2}$} parameter varies as sin(2$\theta$), where $\theta$ is the angle that the vector connecting the binaries makes with the horizontal. This makes it possible to predict $\theta$ via 
\begin{equation}
\theta = \frac{tan^{-1} ( \frac{e_{2}}{e_{1}} )}{2}
\end{equation}

In our test dataset, the median difference between the PTF-predicted orientation and the observed Robo-AO orientation is 2.1$^{\circ}$. 

\begin{figure}
\includegraphics[width=\columnwidth]{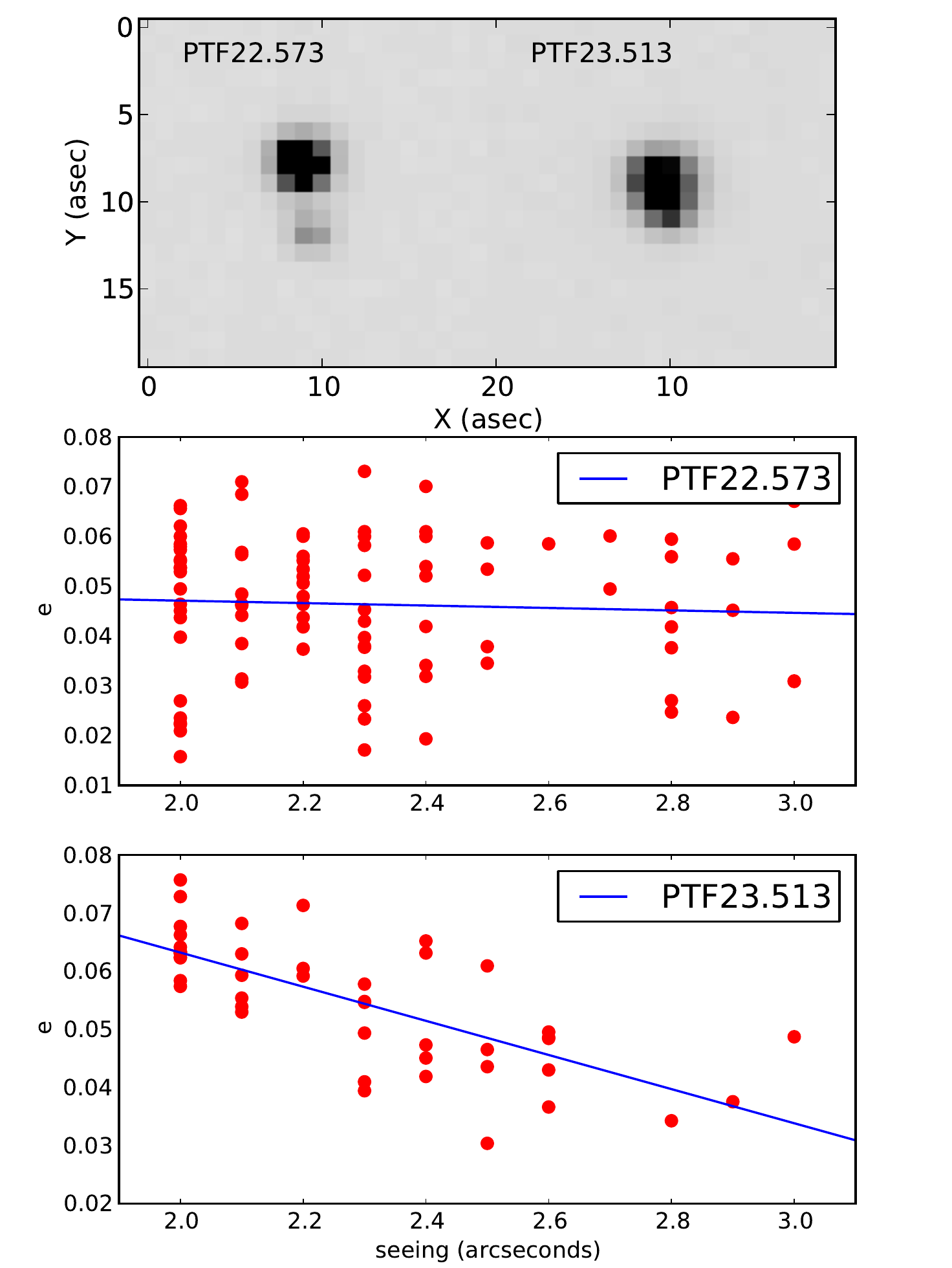}
\caption{Comparison of the seeing dependence of the ellipticity of two binary systems which have a similar measured ellipticity at a common reference FWHM. The images of the two binary systems are from the PTF survey. The ellipticity of the binary whose members are closer together is much more sensitive to the FWHM, while the ellipticity of the binary whose members are further apart is almost constant with changes in FWHM. This difference in seeing dependence allows us to resolve the seeing/contrast degeneracy of the ellipticity.}
\label{fig:seeing_dep_individual}
\end{figure}

\begin{figure*}
\centering
\subfigure{\includegraphics[width=0.33\textwidth]{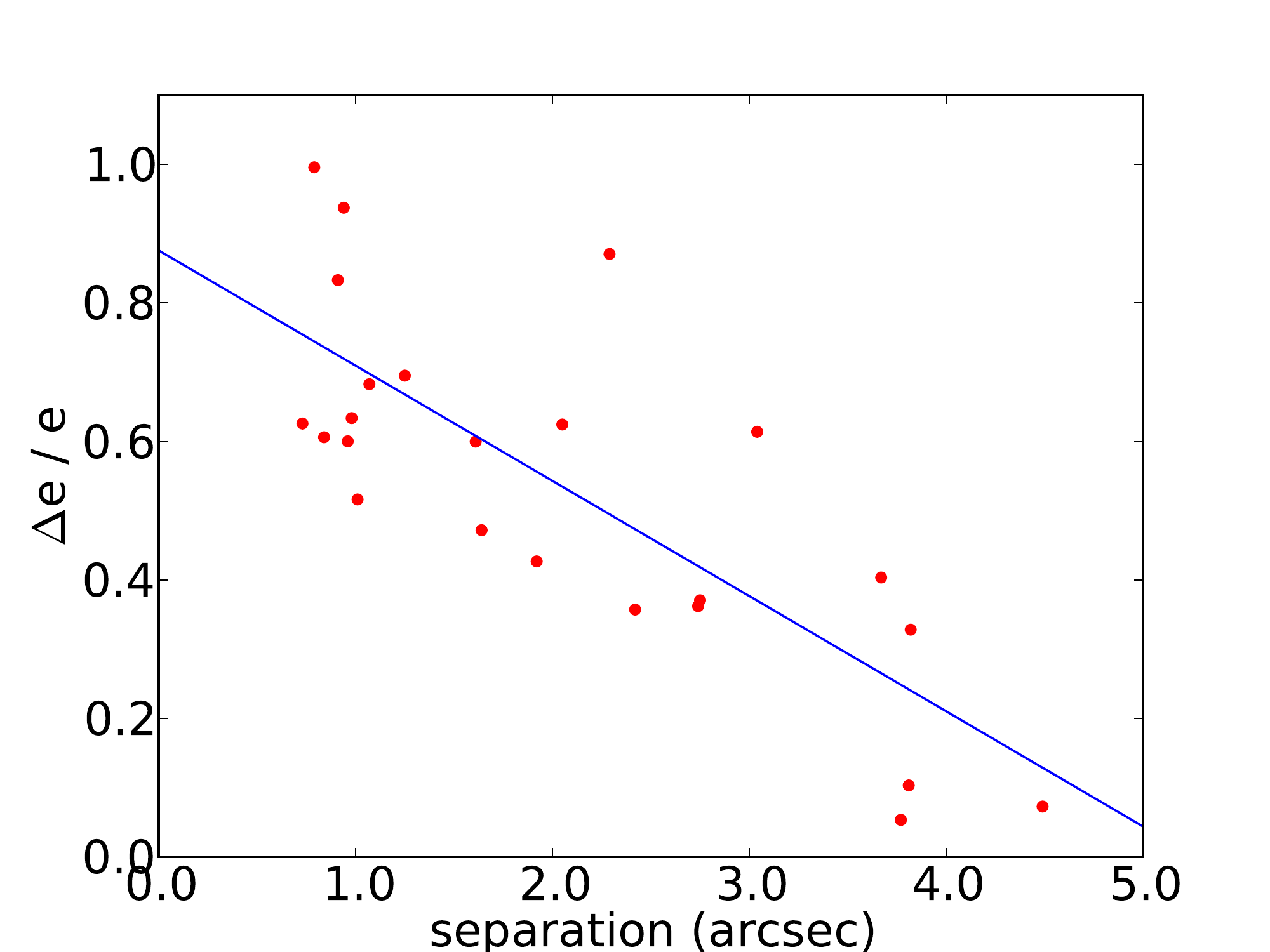}}
\subfigure{\includegraphics[width=0.33\textwidth]{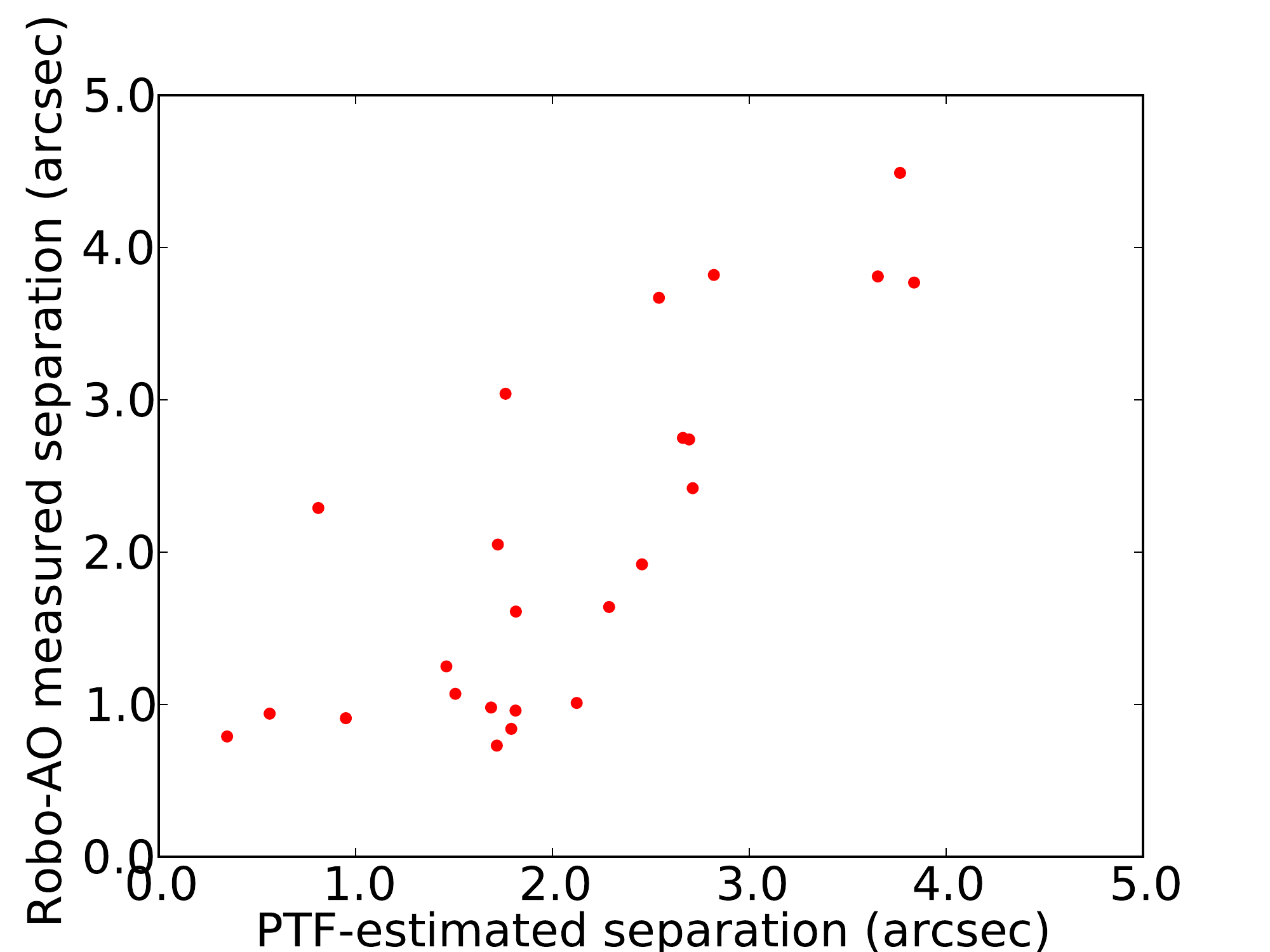}}
\subfigure{\includegraphics[width=0.33\textwidth]{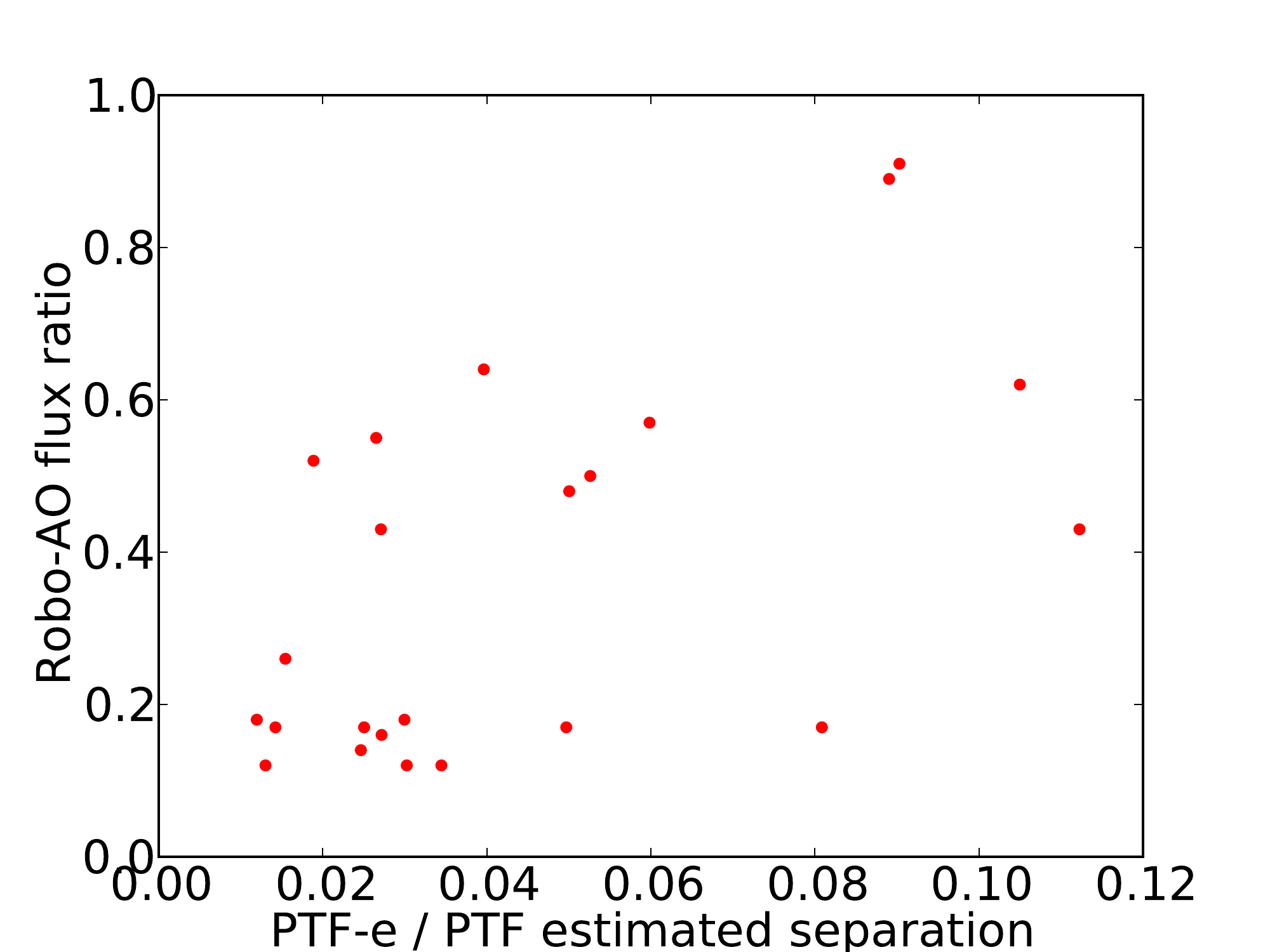}}
\caption{\textit{Left:} the change in ellipticity with seeing (as a fraction of the magnitude of ellipticity) for the binaries we imaged with Robo-AO. For larger separations, the ellipticity becomes less responsive to the FWHM. \textit{Middle:} comparison between the separations estimated from PTF data alone and the Robo-AO measured separations; the correlation shows separation measurement at $\sim$25\% precision. \textit{Right:} comparison between the PTF ellipticity divided by PTF separation, and the Robo-AO measured flux ratio. Although the correlation is weak, there is a significant excess of points towards the lower left, suggesting that the contrast ratio of the binaries can be weakly constrained using PTF data alone.}
\label{fig:seeing_dep}
\end{figure*}

\subsection{Separations and contrast ratios}
The ellipticity of a close binary increases when the flux ratio of the stars in the binary system increases, or alternatively when the separation between the binaries increases (Figure \ref{fig:crit}). Because of this degeneracy between separation and contrast, it is not possible to estimate the binary parameters by examining solely the magnitude of the ellipticity. However, we can differentiate between the two cases based on how the ellipticity varies with seeing.

When the members of a binary are very close together, the PSF of their combined light will only be noticeably elliptical for smaller FWHM values. As the FWHM increases, the PSF will quickly start to resemble a circular PSF. On the other hand, when the two members of a binary are separated by a large angular distance, their combined PSF will remain elliptical for much larger FWHMs than the close binaries. Thus, if the magnitude of the  ellipticity of a binary is relatively constant with seeing, the binary has a wide separation, and if the ellipticity decreases quickly with seeing, the binary members must be close together. 

For example, two of the binaries from Figure \ref{robo_ao}, PTF22.573 and PTF23.513, have nearly identical ellipticities at the chosen reference FWHM of 2.5 arcseconds (as measured from the PTF images). The high-resolution Robo-AO images reveal that these two binaries differ in contrast and separation -- PTF22.573 has greater angular separation and lower contrast ratio than PTF23.513. Figure \ref{fig:seeing_dep_individual} shows the seeing dependence of the ellipticity of both of these binaries. It is apparent than the closer, higher contrast binary has an ellipticity which is much more sensitive to the FWHM, as expected. This difference in response to changes in seeing allows us to resolve the degeneracy in the ellipticity measurement between separation and contrast.

For the binaries we found with Robo-AO, we examined how the measured ellipticity in the PTF images varied with seeing (Figure \ref{fig:seeing_dep}). We derived a simple linear fit of the ellipticity as a function of seeing for each of the binaries. We evaluated these fits at a reference FWHM of 2.5 arcseconds, and we parametrized the response of the ellipticity to seeing with the fractional decrease in seeing per arcsecond (\emph{$\Delta$e / e}, where \emph{$\Delta$e} is the magnitude of the slope of the linear fit of ellipticity with respect to seeing, and \emph{e} is the ellipticity measured at the reference radius of 2.5 arcseconds). As can be seen in Figure \ref{fig:seeing_dep}, the binaries with widest angular separations have close to zero variance of ellipticity with seeing. Using the measured relation to predict the Robo-AO measured separation on the basis of PTF data alone confirms BinaryFinder can measure the binary separations to $\sim$25\% precision. 

The measured ellipticity is a function of flux ratio and separation only, and so with a measured separation the flux ratio of the binary can also be estimated. Also in Figure \ref{fig:seeing_dep}, we use the estimated separations to constrain the flux ratios of the systems, assuming a simple linear relationship between separation and ellipticity independent of contrast ratio (which is approximately correct; see Section \ref{sec:testing}). The resulting distribution is (weakly) correlated with the Robo-AO measured flux ratios; although there are outlier points, a low value of ellipticity-over-separation strongly suggests a high flux ratio between the components of the binary systems.

\section{Discussion, Applications and Conclusions}
\label{sec:concs}

\subsection{False positives}

False positives can be generated in our algorithm by instrumental effects that mimic PSF ellipticity, or extended astrophysical objects which are not physically-associated binaries.

The careful attention to input data quality and PSF anisotropy removal results in a low level of instrumental false positives. Of the 44 targets observed with Robo-AO, 25 had ellipticities greater than our blend detection criterion of 0.02 ellipticity. All 25 of these targets were confirmed to be binary systems, suggesting a false-positive rate of less than 5\% level. 

Proper motion sample cuts can ensure that the BinaryFinder targets are relatively nearby stars rather than galaxies and other extended extragalactic objects. We note, however, that if the proper motion cuts are relaxed, the technique detailed in this paper can detect and measure the characteristics of any objects that lead to an elliptically-shaped image. Beyond binary detection, the algorithm could thus be used to search for and statistically characterize faint barely-extended galaxies, strong gravitational lenses and other extragalactic objects.

\subsection{Very-large-sample binary detection}
We have demonstrated that BinaryFinder can detect binaries down to $\approx$1/5 of the seeing limit in wide-field survey data, and directly measure their position angles, separations and contrast ratios. When applied to the synoptic sky survey data currently being collected in great quantities by synoptic sky surveys like PTF and Pan-Starrs, we will be able to perform a search for binary systems over very large sky areas.

At the time of writing PTF has imaged 1200 fields with sufficient numbers of epochs for full BinaryFinder binary detection and measurement of position angles and separations. Averaging across the sky, each PTF field contains $\approx$10,000 objects which are bright enough ($\rm m_R<18$) for the detection of companions at separations down to $\sim$1/5 of the seeing limit. In typical fields, we find that approximately 35\% of those objects have high enough proper motions to confirm that they are nearby stars, and $\approx$11\% of those are detected as $<$ 2 arcsecond separation binaries by our algorithm (either binaries or blended unassociated foreground/background stars). Estimating the background blend probability from the stellar density in our images, we estimate that only $\approx$1\% of those targets will be background blends. 

In total, a full PTF-dataset BinaryFinder search could target $\sim$12,000,000 stars; we would expect to detect $\sim$450,000 binary systems with detectable proper motions and a high confidence of physical association. A binary sample of that size will allow unprecedentedly detailed statistical analysis of the binarity fraction as a function of stellar mass, age, metallicity and galactic population.

PTF and similar surveys operate in multiple filters. Multi-color imaging ellipticity measurements, if available, would scale according to the relative colors of the binary components, and could be used to reduce the false-positive probability or further constrain the nature of detected systems.

\subsection{Orbital Measurements for an Extremely Large Binary Sample}
Many current and planned synoptic sky surveys image the same areas of sky repeatedly over many years. With the ability to directly measure position angle, separation and contrast ratio of individual binaries, in many cases sufficient data will be collected to follow the motion of binary systems with time. Since this can be done over large areas and very large samples, BinaryFinder could be used to obtain orbital measurements (and thus mass constraints) for a very large sample of binaries.

\subsection{Summary and conclusions}

We have shown that BinaryFinder is capable of detecting and characterizing close binary systems in wide-field synoptic survey data. Using Robo-AO, we have confirmed the PTF direct detection and separation measurement of binaries as small as 0.4 arcsec, or less than 1/5 of the PTF FWHM.  Images from the PTF camera have a pixel sampling of 1.01 arcseconds and a median FWHM of $\approx$2.0 arcseconds. Because our algorithm requires well-sampled images, we restricted its operation to PTF images with 2--3 arcsec FWHMs. New synoptic sky surveys with smaller pixels and smaller PSFs will enable the detection of much closer binary systems; for well-sampled images the detection limit scales with the PSF FWHM (assuming similar systematic noise performance). Using BinaryFinder, a survey with improved sampling such as LSST is likely to be capable of the detection, characterization and orbital motion measurement of millions of multiple systems with separations as close as 0.1 arcseconds.

\section{Acknowledgements}
We thank Yanqin Wu for very useful discussions, and the Palomar Observatory staff for their superb support of Robo-AO and PTF operations. ET participated in the Summer Undergraduate Research Program (SURP) at the Dunlap Institute for Astronomy \& Astrophysics, University of Toronto. NML is a Dunlap Fellow at the Dunlap Institute for Astronomy \& Astrophysics, University of Toronto.  The Dunlap Institute is funded through an endowment established by the David Dunlap family and the University of Toronto. The Robo-AO system is supported by collaborating partner institutions, the California Institute of Technology and the Inter-University Centre for Astronomy and Astrophysics, and by the National Science Foundation under Grant Nos. AST-0906060 and AST-0960343, by a grant from the Mt. Cuba Astronomical Foundation, and by a gift from Samuel Oschin. Observations were obtained with the Samuel Oschin Telescope at the Palomar Observatory as part of the Palomar Transient Factory project, a scientific collaboration between the California Institute of Technology, Columbia University, Las Cumbres Observatory, the Lawrence Berkeley National Laboratory, the National Energy Research Scientific Computing Center, the University of Oxford, and the Weizmann Institute of Science.  ALK was supported by NASA through Hubble Fellowship grant 51257.01 awarded by STScI, which is operated by AURA, Inc., for ANSA, under contract NAS 5-26555.

\bibliographystyle{apj}
\bibliography{refs}

\end{document}